# DiPCo - Dinner Party Corpus

*Maarten Van Segbroeck[1], Ahmed Zaid[1], Ksenia Kutsenko[1], Cirenia Huerta[1], Tinh Nguyen[1], Xuewen Luo[1], Björn Hoffmeister[1,\*],*
*Jan Trmal[2], Maurizio Omologo[3], Roland Maas[1]*

[1]Amazon
[2]Center for Language and Speech Processing, Johns Hopkins University, Baltimore, USA
[3]ICT-irst, Fondazione Bruno Kessler, Trento, Italy

`segbrm, ahmemu, ksenik, cireni, ngtinh, xuewl, bjornh, rmaas@amazon.com, jtrmal@gmail.com, omologo@fbk.eu`

*Abstract*—We present a speech data corpus that simulates a "dinner party" scenario taking place in an everyday home environment. The corpus was created by recording multiple groups of four Amazon employee volunteers having a natural conversation in English around a dining table. The participants were recorded by a single-channel close-talk microphone and by five far-field 7-microphone array devices positioned at different locations in the recording room. The dataset contains the audio recordings and human labeled transcripts of a total of 10 sessions with a duration between 15 and 45 minutes. The corpus was created to advance in the field of noise robust and distant speech processing and is intended to serve as a public research and benchmarking data set.

## 1 INTRODUCTION

The availability of speech corpora with high-quality transcribed audio recordings plays an important role to progress in the field of speech processing and automatic speech recognition (ASR). These corpora are essential for speech researchers to objectively evaluate and benchmark speech processing algorithms. In this paper, we present the Dinner Party corpus, a speech database that replicates the scenario where a group of people are having an interactive conversation while having dinner in a simulated home environment. The corpus has been designed with the objective to accelerate research in a wide variety of challenging speech processing tasks in near- and far-field acoustic conditions, such as noise robust and conversational speech recognition, speaker identification and speaker separation.

The corpus consists of multiple sessions recorded in the same room over multiple days and with different groups of participants. Each session contains the conversational speech recordings in English of four volunteering participants who are seated around a dining table. Each participant has been recorded by a close-talk microphone and five far-field array microphone devices placed at various location in the room. The close-talk recordings of all speakers were manually transcribed and sentence boundaries were provided. The microphone recordings per session were all time-synchronized. Section 2 describes the corpus in more detail. Baseline ASR results using the Kaldi toolkit are provided in section 3. Availability information is given in section 4.

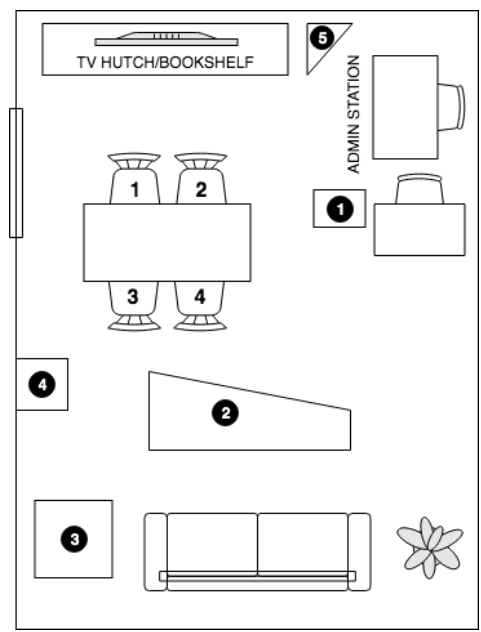

*Figure 1: Layout of the room in which the sessions were recorded.*

## 2 DINNER PARTY CORPUS

### 2.1 Scenario

The corpus contains the collection of 10 sessions in which 4 persons have a natural conversation over dinner. At the beginning of each session, participants were getting food at the buffet and then moved to the dining table. In each session, music playback started at a given time mark (see Table 3). All sessions were recorded in the same room. Figure 1 shows the floor plan and layout of the room. The room dimensions are as follows: length of 6.4 meters (21 feet), width of 4 meters (13 feet) and height of 2.75 meters (9 feet). The room was equipped with 5 far-field microphone array devices that were placed at the following positions: side table (device 1), coffee table (2), side table (3), shelf (4) and on top of 4-tier shelf (5). The four participants were seated at the dining table at the positions indicated in Figure 1. The distance of each speaker to the 5 devices is shown in Table 1.

---

| participant | device | | | | |
|---|---|---|---|---|---|
| | 1 | 2 | 3 | 4 | 5 |
| 1 | 1600 | 2240 | 3825 | 2900 | 1760 |
| 2 | 1990 | 2130 | 3950 | 3100 | 1760 |
| 3 | 1820 | 1520 | 2900 | 2030 | 2780 |
| 4 | 1300 | 1120 | 3100 | 2520 | 2820 |

*Table 1: Distant measured between the participants and the microphone array devices (in mm).*

### 2.2 Data collection

All participants were simultaneously recorded with single-channel close-talking (headsets) and far-field microphone devices. The far-field devices were equipped with a microphone array consisting of 7 microphone channels. The 7-microphone array was configured as illustrated in Figure 2. The 6 microphones were uniformly placed on the perimeter of a circle of radius 35 millimeters, the 7th microphone was placed at the center of the circle. All microphone recordings are provided in 16-bit WAV file format with 16 kHz sampling frequency and were obtained by downsampling the original 45 kHz audio recordings. The name convention used for the files was adopted from [1] and is described in the corpus description file that is included with the corpus data [2]. The total number of microphones per session is 39 (4x close-talk microphones + 5x7 far-field microphones) and are all time-synchronized.

### 2.3 Corpus design

The Dinner Party corpus was designed in a similar fashion as the dataset of [1]. It consists of 10 recorded sessions and has a total of 32 unique speakers. All speakers are adults with an age range between 22 and 62. The male/female ratio is 19/13 and 15 participants were non-native U.S. English speakers. The total number of the transcribed sentences is 7081. The corpus was divided in a development and evaluation set as shown in Table 2.

| Dataset | Sessions | Hours hh:mm | #Utts |
|---|---|---|---|
| Dev | S02, S04, S05, S09, S10 | 02:43 | 3673 |
| Eval | S01, S03, S06, S07, S08 | 02:36 | 3408 |

*Table 2: Corpus overview.*

The sessions have a duration ranging between 15 and 47 minutes as illustrated in Table 3. Sessions were assigned to the development and evaluation set to balance the total duration, the amount of music playback and number of participants. There is no speaker overlap between the development and evaluation sets.

Note: the close-talk microphone recording of participant P13 of session S04 is noisier than others due to a microphone issue. Since all other recordings and transcriptions are good, we decided to keep the session in full in the corpus.

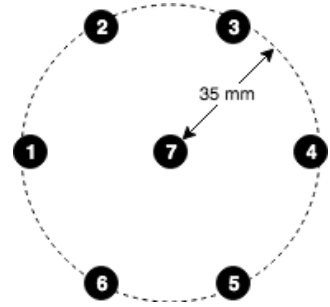

*Figure 2: Configuration of the 7-microphone array.*

### 2.4 Annotations

For each speaker, the transcriptions were obtained by listening to the corresponding close-talk recordings. To determine who the main speaker is of the recording, i.e. the person who is recorded by the close-talk microphone, the transcriber listened to about 15 seconds of audio in which all participants identified themselves, e.g. by saying "speaker 3". For each close-talk recording, the transcriber was asked to only transcribe the main speaker and ignore all audio segments in which the other participants are talking.

| Session | Participants | Hours hh:mm | #Utts | Music hh:mm:ss |
|---|---|---|---|---|
| S02 | **P05**, **P06**, **P07**, P08 | 00:30 | 448 | 00:19:30 |
| S04 | **P13**, P14, **P15**, P16 | 00:45 | 1284 | 00:23:25 |
| S05 | **P17**, **P18**, **P19**, P20 | 00:45 | 1010 | 00:31:15 |
| S09 | P29, **P30**, P31, **P32** | 00:22 | 499 | 00:12:18 |
| S10 | P29, **P30**, P31, **P32** | 00:20 | 432 | 00:07:10 |
| S01 | P01, **P02**, **P03**, P04 | 00:47 | 903 | 00:38:52 |
| S03 | **P09**, **P10**, **P11**, P12 | 00:46 | 1128 | 00:33:45 |
| S06 | **P21**, P22, **P23**, P24 | 00:20 | 462 | 00:06:17 |
| S07 | **P21**, P22, **P23**, P24 | 00:26 | 581 | 00:10:05 |
| S08 | **P25**, P26, P27, P28 | 00:15 | 334 | 00:01:02 |

*Table 3: Session overview. Male participants are in bold and native U.S. English speakers are underscored.*

To determine utterance segment boundaries, the transcriber was requested to search for logical time stamps, such as at the beginning of a new sentence. Long instances of continuous speech were asked to break into segments of up to 10 seconds, with a maximum allowance up to 15 seconds if necessary. If possible, the allowed segment length should give enough context to discern the speaker's emotional state. The transcripts also include the following tags:

- [noise] noise made by the speaker (coughing, lip smacking, clearing throat, breathing, etc.)
- [unintelligible] speech was not well understood by transcriber
- [laugh] participant laughing

Transcript files are provided per session in JSON format and contain the utterance transcriptions of all participants of the session.

## 3 BASELINE ASR RESULTS

Baseline word error rate (WER) numbers were generated by using the Kaldi baseline (`egs/chime5/s5`) [2, 3] as the acoustic model (AM) source on which model adaptation was performed. The AM was a Time Delay Neural Network and Factored (TDNN-F) deep neural network [4] with 15 layers, the dimension of each layer was 1536 and the bottleneck dimension was 160.

In a leave-one-out fashion, the Dev set was split by sessions to create five adaptation sets, each containing four sessions and with the fifth session used for verification purposes. This leave-one-out cross-validation process was set up to robustly infer the following meta-parameters: language model interpolation weight, number of iterations to use for adaptation, language model weight and word-insertion penalty. The adaptation was implemented as training with reduced learning rate (factor 10x smaller than the original learning rate). Note that we did not experiment with freezing only some of the layers, hence all network layers were retrained. After adaptation, the verification session was decoded.

For the language model, we interpolated the text data from the adaptation set with the Cantab-TEDLIUM v1.1 [5] language model. The interpolation weight was typically around 0.5 and the reduction of perplexity was more than 50% absolute (for example for session S02, the perplexity went from 390 to 190). We created two rescoring LM models for the final experiments. The first was obtained by interpolation of the LM with the Cantab-TEDLIUM LM named *LM3*. The second model is an interpolated 4-gram unpruned LM created from the Cantab-TEDLIUM LM named *LM4*. The Cantab-TEDLIUM lexicon was also used as the baseline lexicon. We trained a Phonetisaurus G2P system to generate pronunciations for words that were OOVs w.r.t the Cantab-TEDLIUM lexicon.

Table 4 shows results on the verification sets and overall Dev performance. The overall Dev performance was obtained by combining all the verification sets decoding outputs, which yields a decoded DEV set, where each utterance was decoded by a model not trained nor conditioned on that given utterance. Results are shown for close-talk and far-field recordings of all sessions and were averaged over all speakers/devices. The numbers suggest that the best strategy is to not adapt for close-talk microphone recordings. For far-field recordings, the best results were obtained by using 4 adaptation iterations.

The leave-one-out strategy provides us a fairly unbiased estimate about the best set of meta parameters used for the Eval set. Table 5 shows the results on the Eval set using the best set of meta parameters for a model that has been adapted using all Dev sessions. The best WER is obtained by rescoring using the interpolated 4-gram unpruned LM from Cantab-TEDLIUM.

## 4 AVAILABILITY

The Dinner Party corpus (DiPCo) can be downloaded in compressed tar.gz format [6]. The corpus is made available under the CDLA-Permissive license [7].

| Verification session | no adaptation | | after 4 iterations | |
|---|---|---|---|---|
| | *CT* | *FF* | *CT* | *FF* |
| S02 | 46.52 | 83.84 | 48.63 | 73.18 |
| S04 | 47.70 | 91.43 | 48.78 | 84.20 |
| S05 | 59.91 | 91.32 | 61.98 | 86.87 |
| S09 | 35.67 | 85.08 | 36.85 | 74.79 |
| S10 | 39.96 | 86.97 | 39.78 | 69.89 |
| overall | ***48.21*** | 88.60 | 49.54 | ***79.70*** |

Table 4: WER (%) results on the close-talk (CT) and far-field (FF) recordings of the Dev set session. Results shown are before and after adapting the AM on the Dev set sessions while leaving one session out for verification purposes.

| Eval session | no rescoring | | LM3 rescoring | | LM4 rescoring | |
|---|---|---|---|---|---|---|
| | *CT* | *FF* | *CT* | *FF* | *CT* | *FF* |
| S01 | 54.59 | 77.59 | 50.06 | 76.08 | 49.68 | 75.92 |
| S03 | 45.28 | 76.85 | 40.84 | 75.48 | 40.49 | 75.26 |
| S06 | 44.83 | 79.87 | 39.93 | 78.12 | 39.24 | 77.99 |
| S07 | 46.83 | 76.45 | 41.98 | 74.01 | 41.30 | 73.85 |
| S08 | 44.86 | 84.71 | 40.45 | 84.74 | 39.78 | 84.57 |
| averaged | 47.28 | 79.09 | 42.65 | 77.69 | ***42.10*** | ***77.52*** |

Table 5: Final WER (%) results on the close-talk (CT) and far-field (FF) recordings of the Eval set session. Results shown with the AM retrained on all Dev set sessions, before and after rescoring with LM3 and LM4.

## 5 ACKNOWLEDGMENT